\documentclass[showpacs,preprintnumbers,twocolumn]{revtex4-1}

\usepackage{amssymb}

\usepackage{amsfonts}

\usepackage{latexsym}

\usepackage{dcolumn}

\usepackage{amsmath}

\usepackage{graphicx}

\usepackage{psfrag,pstricks}

\begin{document}

\title{Microwave Photodetection with Electro-Opto-Mechanical Systems}

\author{Shabir Barzanjeh$^{1}$}
\author{M. C. de Oliveira$^{2}$}
\author{Stefano Pirandola$^{3}$}
\affiliation{$^{1}$Institute for Quantum Information, RWTH Aachen
University, 52056 Aachen, Germany} 
\affiliation{$^{2}$Instituto de
F\'\i sica Gleb Wataghin, Universidade Estadual de Campinas,
13083-970, Campinas, SP, Brazil,} \affiliation{$^{3}$Department of
Computer Science, University of York, York YO10 5GH, United
Kingdom}
\date{\today}

\begin{abstract}
While detection of optical photons is today achieved with very
high efficiencies, the detection of microwave fields at the photon
level still poses non-trivial experimental challenges. In this
Letter we propose a model of microwave photodetector which is
based on the use of an electro-opto-mechanical system. Using state
of the art technology, we show how single microwave photons can
efficiently be converted into optical photons which are then
measured by standard optics. The overall quantum efficiency of our
microwave detector tends to one for achievable values of the
optical and microwave cooperativity parameters. Our scheme could
be used for sensing and scanning very faint microwave fields, as
they may occur in radio astronomy, satellite and deep space
communications.
\end{abstract}


\maketitle \textit{Introduction.} Detection of photons is required
for many fundamental and technological purposes. It is
particularly relevant in the quantum regime, where the detection
of few photons must be resolved, for both photocounting and access
to the statistics of radiation field
states~\cite{Milburn,Scully,Yamamoto}.
In the optical, near optical, and infrared domain, photodetection
it is readily available through semiconductor technologies, where
single photon resolution has been achieved. However microwave
frequencies of the radiation are not directly reachable due to
technological limitations. Several applications in cavity quantum
electrodynamics (QED) have employed Rydberg atoms population
detection as an indirect way of determining the coherence of
Fabry-Perrot microwave cavity fields, but inevitably the atoms
detection induce back-action over the field states~\cite{Haroche}.

Nowadays a large effort has been devoted to circuit QED, where
transmission line resonators are coupled to superconducting
devices, known as  \lq artificial atoms\rq. Similarly to the
Fabry-Perot cavities, indirect detection of charge or flux
populations of the artificial atoms is employed as a way to infer
physical properties of the radiation field~\cite{Devoret}.
Unfortunately, the access to the statistical properties is only
available when one can also access the number of microwave
photons. Recent proposals for microwave photodetection using
circuit QED have been demonstrated with limited quantum efficiency
considering nowadays available superconducting
technology~\cite{Romero,Peropadre,Poudel,Fan}.

Moreover it would be interesting to have a photodetector able to
work on demand in many different experimental situations and not
restricted to a specific setup, similarly to the photodetectors
used in the optical domain. One possible strategy to achieve this
goal and circumvent both the limited detection efficiency and the
problem of detection backaction is to perform a conversion of
microwave photons into optical ones, where efficient
photon-detection is readily available. This is indeed the strategy
suggested in our Letter.

We propose a microwave photodetector which is based on the
efficient conversion of microwave photons into optical. This is
realized by using an electro-opto-mechanical system where a
mechanical oscillator induces strong coupling between a microwave
and an optical resonator field. Under specific detuning of both
the microwave and optical pumping fields, the average number of
optical photons in the signal output is directly proportional to
the mean number of microwave photons in the signal input. The
proportionality parameter is an effective quantum efficiency which
depends on optical and microwave cooperativity parameters
associated with the system.

Adopting achievable experimental parameters for the
electro-opto-mechanical transducer, we show that the efficiency of
the microwave-optical conversion can be extremely high. Combining
this feature with the use of optical photodetection on the
converted optical photons, we can achieve an overall quantum
efficiency which is close to 100\%. In our proposal there is no
specific requirement for the input microwave signal, so that the
setup can be used for microwave photon detection under several
circumstances. In particular, it could be used for sensing faint
microwave fields at the photon or sub-photon level, with potential
applications which include the environmental scanning of
electrical circuits, radio astronomy, satellite and deep space
communications.
\begin{figure}[h]
\centering
\includegraphics[width=3.2in]{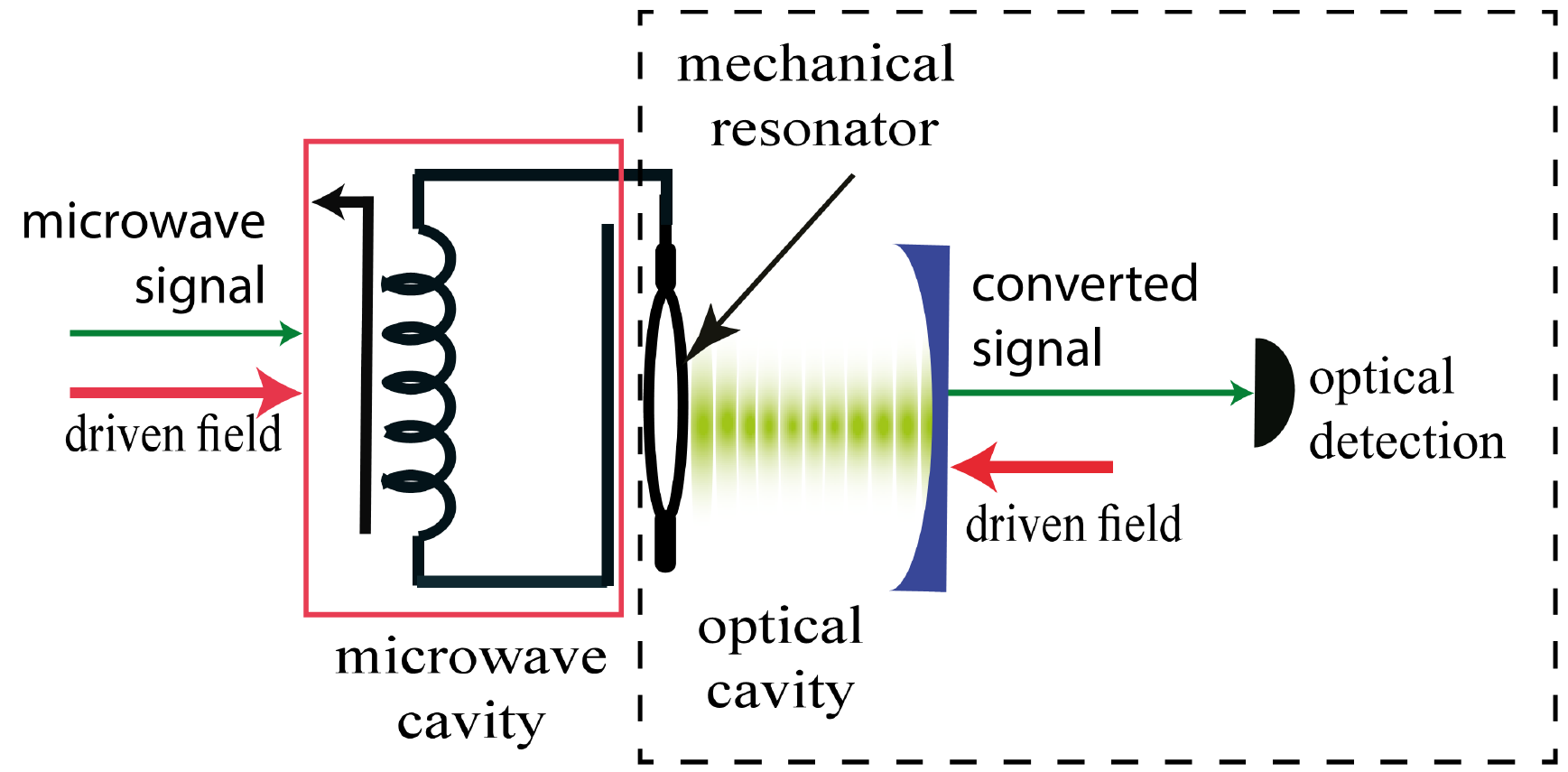}
\caption{Schematic description of our microwave photodetector
based on a electro-opto-mechanical converter. The microwave signal
photons are first converted into optical photons and then measured
by a standard optical photodetector.} \label{fig1}
\end{figure}

\textit{System.}  As sketched in Fig. \ref{fig1}, the system
consists of two microwave and optical cavity modes, coupled to a
single mechanical resonator~ \cite{shabir1,lenh1, bochmann}, with
frequency $ \omega_M $ and damping rate $ \gamma_M $. The
microwave~(optical) cavity works at frequency $
\omega_{\mathrm{w}}(\omega_\mathrm{o}) $ with the total cavity
decay rate being $ \kappa_j
=\kappa_j^{\mathrm{ext}}+\kappa_j^{\mathrm{int}}$. We include the
intrinsic losses with rates $\kappa_j^{\mathrm{int}}$
$~(j=\mathrm{w},\,\mathrm{o})$ for both optical and microwave
cavity modes, while $ \kappa_j^{\mathrm{ext}} $ denote the
coupling rates of the respective input ports.

The optical and microwave cavity photons interact with the
phononic modes of the mechanical via radiation pressure forces.
This interaction is described by the following
Hamiltonian~\cite{shabir1,shabir2}
\begin{eqnarray}\label{lan1}
\hat H&&=\hbar\omega_M \hat b^{\dagger}\hat b+\hbar\sum_{j=\mathrm{w},o}\Big[\Delta_j+g_j(\hat b+\hat b^{\dagger})\Big]\hat a_j^{\dagger}\hat a_j\nonumber\\&&+i\hbar\sum_{j=\mathrm{w},o} E_j(\hat a_j^{\dagger}-\hat a_j),
\end{eqnarray}
where $\hat b$ is the annihilation operator for mechanical
resonator, $\hat a_j$ is the annihilation operator for the cavity
$ j=\mathrm{w},o $, and $ g_j $ is the single photon
electro-opto-mechanical coupling rate between cavity $ j $ and
mechanical resonator. Here we also assume that the microwave and
optical cavities are driven at the frequencies
$\omega_{d,j}=\omega_{j}- \Delta_j$, where $ \Delta_j $ is the
detuning of the cavity $j=\mathrm{w},o$ and $ E_j $ are the
amplitudes of the driven pumps.

We can linearize the previous Hamiltonian by expanding the cavity
modes around the steady state field amplitudes in each resonator.
This is equivalent to set $ \hat c_j=\hat a_j-\sqrt{N_j}$, where
$N_{j}=|E_j|^2/(\kappa_j^2+\Delta_j^2)\gg 1$ are the strength of
the pumps, expressed in terms of mean number of cavity photons
induced by the microwave and optical pumps~\cite{shabir2, lenh}.
The effective Hamiltonian of the system, in an interaction picture
with respect to the free Hamiltonian, is therefore given by
 \begin{equation}\label{hameff}
\hat H=\hbar G_\mathrm{o}(\hat c_\mathrm{o} \hat b^{\dagger}+\hat
b \hat c_\mathrm{o}^{\dagger})+\hbar G_{\mathrm{w}}(\hat
c_{\mathrm{w}} \hat b^{\dagger}+\hat b \hat
c_{\mathrm{w}}^{\dagger}),
\end{equation}
where $ G_j=g_j\sqrt{N_j} $ are many-photon optomechanical
couplings.

In the Hamiltonian of Eq.~(\ref{hameff}) we have set the cavity
detunings to be $\Delta_{\mathrm{w}}=\Delta_\mathrm{o}=\omega_M $
and assumed the regime of fast mechanical oscillations, so that we
are in the resolved sideband regime for both cavities, with red
sideband driving for both microwave and optical cavities. In this
regime we have neglected the fast oscillating terms proportional
to $ \pm 2 \omega_M $. The first term in Eq.~(\ref{hameff})
describes a beam-splitter like interaction between the mechanical
resonator and propagating optical fields in the fibre at a rate $
F_\mathrm{o}:= G_{\mathrm{o}}^2/\kappa_\mathrm{o}$. Similarly the
second term describes the coherent exchange of the excitations
between the mechanical resonator and the cavity microwave field at
a rate $ F_\mathrm{w}:= G_{\mathrm{w}}^2/\kappa_\mathrm{w}$. Note
that the process of the exchange of excitations is coherent, as
long as $ \gamma_Mk_BT/(\hbar \omega_M)< F_j $~($
j=\mathrm{w},\,\mathrm{o} $)~\cite{lenh,hofer}, where $ k_B $ is
Boltzmann constant, and $ T $ is the temperature of the
electro-opto-mechanical converter.

By using quantum Langevin equations~\cite{langevin} and standard
input-output theory~\cite{Milburn}, the output variable of the
optical cavity $\hat c_{\mathrm{o,out}}=
\sqrt{2\kappa_\mathrm{o}^{\mathrm{ext}}}\hat c_\mathrm{o}-\hat
c_{\mathrm{o,ext}}$ is given by
\begin{eqnarray}
 \hat c_{\mathrm{o,out}}(\omega)=&&-A(\omega)\hat c_{\mathrm{o,ext}}-B(\omega)\hat c_{\mathrm{w,ext}}
 -C(\omega) \hat b_{\mathrm{in}}\nonumber \\ &&-D(\omega) \hat c_{\mathrm{w},\mathrm{int}}-E(\omega) \hat
c_{\mathrm{o,int}},\label{cOUT}
\end{eqnarray}
where the coefficients $ A$, $B$, $C$, $D $ and $ E $ satisfy
$|A|^2+|B|^2+|C|^2+|D|^2 +|E|^2=1$ and depend on the cooperativity
terms $\Gamma_j=G_j^2/(\kappa_j \gamma_M)$ (see~\cite{parameters} for details). 

\textit{Pulse conversion.} We assume the incoming microwave pulse
signal is a coherent pulse with non-zero frequency spread~$ W $.
We choose such a coherent pulse to have its center frequency at
the bare resonance frequency of the microwave cavity, i.e., $
\omega_p=\omega_\mathrm{w} $. Therefore, the photon flux per unit
angular frequency (power spectrum) is given by
$\bar{n}_\mathrm{w,ext}(\omega)= \langle \hat
c_{\mathrm{w,ext}}^{\dagger}(\omega) \hat
c_{\mathrm{w,ext}}(\omega)\rangle=|\alpha(\omega)|^2 $, where
\begin{equation} \nonumber \alpha(\omega)=\frac{\alpha_0
\,e^{-(\omega-\omega_p)^2/W^2}}{(2\pi)^{1/4}\sqrt{W/2}},
\end{equation}
with $\alpha_0$ providing the total number of photons in the input
pulse via $ n_\mathrm{p}= \int d\omega \;
\bar{n}_\mathrm{w,ext}(\omega)=|\alpha_0|^2 $. We see that
$\bar{n}_\mathrm{w,ext}(\omega)$ is Gaussian~\cite{RMP} with
standard deviation $W/2$.

From Eq.~(\ref{cOUT}) we can compute the power spectrum of the
field at the output of the optical cavity, i.e.,
$\bar{n}_\mathrm{o,out}(\omega)=\langle \hat c_{\mathrm{o,out}}^{
\dagger}(\omega)\hat c_{\mathrm{o,out}}(\omega)\rangle$. This is
given by
\begin{eqnarray}\nonumber
\bar{n}_\mathrm{o,out}(\omega)=|B(\omega)|^2
\bar{n}_\mathrm{w,ext}(\omega)+N_{\mathrm{noise}}(\omega),
\label{qlessimplify3}
\end{eqnarray}
which depends on the microwave spectrum
$\bar{n}_\mathrm{w,ext}(\omega)$ and
\begin{equation}
N_{\mathrm{noise}}=\Big[|C(\omega)|^2n_b^T
+|D(\omega)|^2n_\mathrm{w}^T+(|A|^2+|E|^2)n_\mathrm{o}^T\Big]\delta(\omega)\nonumber
\end{equation}
corresponding to Markovian noise added by the conversion process.
Here $n_b^T   $ and $n_\mathrm{w}^T$ represent the thermal numbers
of the mechanical and microwave baths, respectively, while
$n_\mathrm{o}^T=[\mathrm{exp}(\hbar\omega_{\mathrm{o}}/k_{B}T)-1]^{-1}\approx0$
is the thermal occupation number for the intracavity optical
field. The mean photon number of the output optical field is equal
to $\bar N_\mathrm{o}=\int d\omega ~\bar{n}_\mathrm{o,out}(\omega)$.
\begin{figure}[ht]
\centering
\includegraphics[width=2.8in]{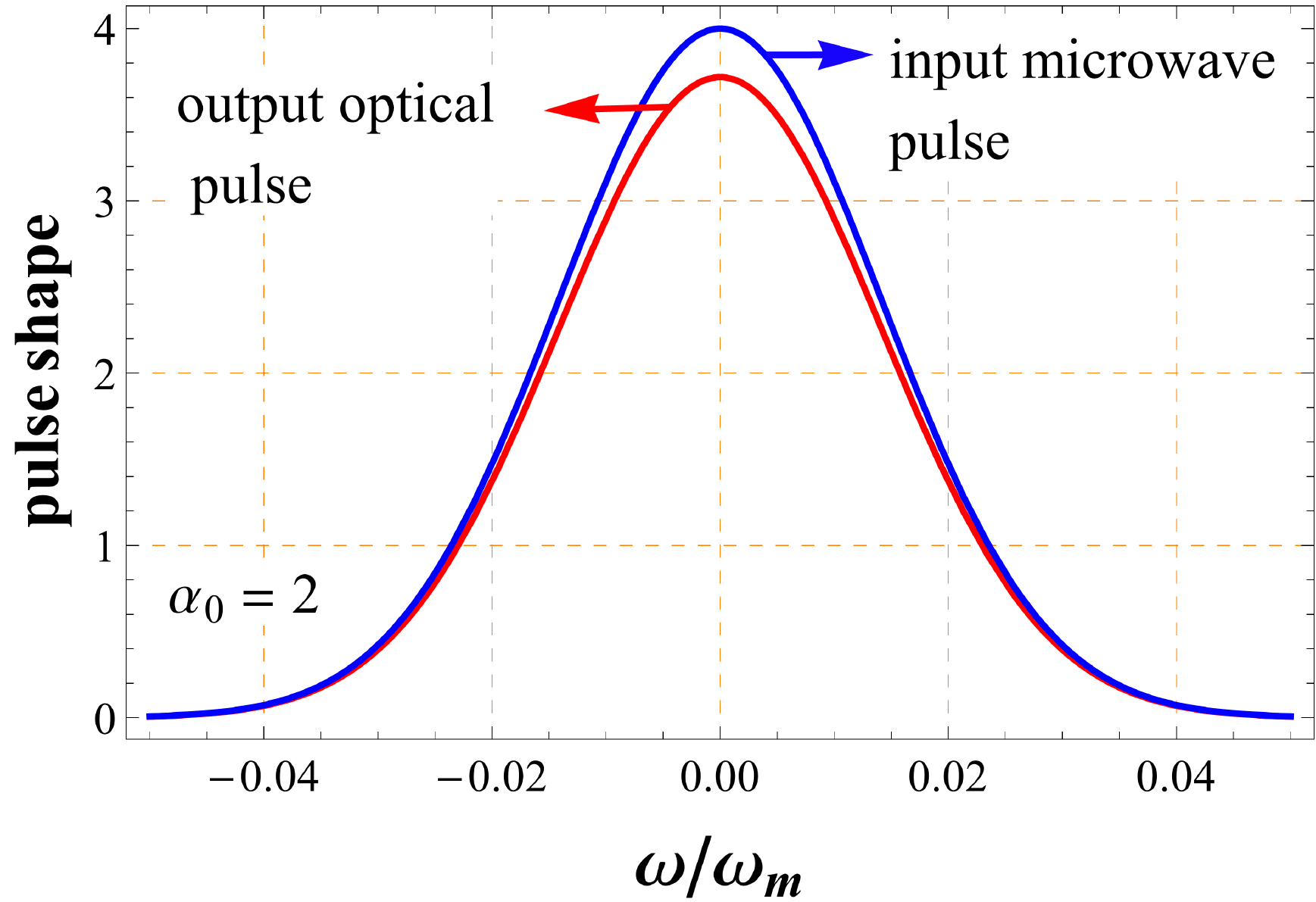}
\caption{The power spectrum of the incoming microwave pulse
$\bar{n}_\mathrm{w,ext}(\omega)$ and that of the converted optical
pulse $\bar{n}_\mathrm{o,out}(\omega)$ are plotted versus the
normalized frequency $\omega/\omega_M $. Here, we have assumed a
mechanical resonator with frequency $\omega_M/2\pi=10$ MHz,
quality factor $Q=36\times 10^4$ and mass $m = 10$ ng, which
interacts with a microwave cavity with
$\omega_{\mathrm{w}}/2\pi=10$ GHz,
$\kappa_{\mathrm{w}}=0.101\omega_M$, driven by a microwave source
with power $P_{\mathrm{w}}=35$ mW.  We have then considered an
optical cavity of length $L=1$ mm and damping rate
$\kappa_c=0.301\omega_M$, which is driven by a laser with
wavelength $\lambda_{0c}=1064$ nm and power $P_c=5mW$. The whole
system is located at the cryogenic temperature $ T=4 $K. We have
assumed the total number of photons in the incoming microwave
pulse is $n_p=4$ and its bandwidth $ W= 1.7 $MHz. This is less
than the bandwidth of the electro-opto-mechanical converter, i.e.,
$ W=0.1W_c$, where $
W_c=|\gamma_M(1+\Gamma_{\mathrm{w}}+\Gamma_{\mathrm{o}})| $.}
\label{fig2}
\end{figure}

In order to see the effect of the converter on the shape of the
incoming pulse, we plot in Fig.~\ref{fig2} both the incoming
(microwave) pulse and converted (optical) pulse with respect to
the normalized frequency $\omega/\omega_M  $. In this figure we
have taken experimental-achievable parameters for the
electro-opto-mechanical converter~\cite{lenh, palo} which is
assumed to operate at cryogenic temperatures ($ T=4 $K). We see
how a faint microwave pulse ($n_p=4$ photons) is successfully
converted into an output optical pulse.

\begin{figure}[h]
\centering
\includegraphics[width=3.2in]{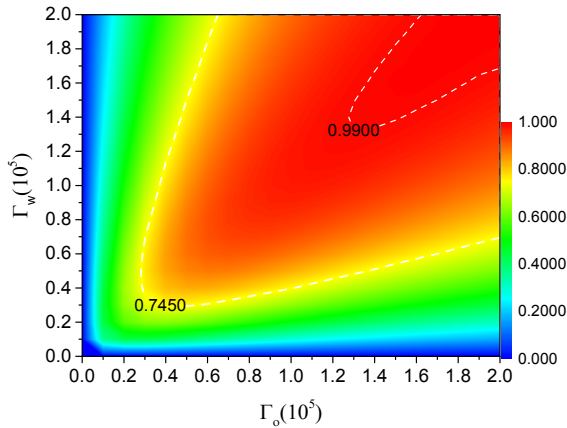}
\caption{Ratio between output optical photons and input microwave
photons $\bar N_\mathrm{o}/n_p$ plotted versus the two
cooperativity parameters $ \Gamma_\mathrm{w} $ and $
\Gamma_\mathrm{o} $. The parameters are the same as
Fig.~\ref{fig2}.} \label{fig3}
\end{figure}

 To analyze the general efficiency of the converter, we consider
the ratio between the output optical photons and the input
microwave photons, i.e., $\bar N_\mathrm{o}/n_p$. This ratio
depends on the values of the two cooperativity parameters
$\Gamma_\mathrm{w} $ and $\Gamma_\mathrm{o}$. As we can see from
Fig.~\ref{fig3}, the larger these parameters are, the better is
the microwave to optical conversion, which rapidly approaches the
ideal conversion rate of 100$\%$.

\textit{Quantum efficiency of the microwave detector.} At the
output of the converter, the optical photons are measured by a
photodetector with high efficiency $\eta$ and large bandwidth (up
to 10GHz, i.e., much larger than $W_\mathrm{c}$). Assuming an
incoming microwave pulse with small bandwidth (so that we can
approximate the Gaussian pulse with a delta function), the mean
number photons which are detected at the optical output is equal
to
\begin{eqnarray}\label{efficency} \bar N_o\simeq
\eta_\mathrm{eff} n_p+\eta N_{\mathrm{thermal}},
\end{eqnarray}
where $ \eta_\mathrm{eff}:= \eta |B(0)|^2 $ is the quantum
efficiency of the microwave detector, and $
N_{\mathrm{thermal}}=|C(0)|^2n_b^T +|D(0)|^2n_\mathrm{w}^T $.
\begin{figure}[h]
\centering
\includegraphics[width=3.2in]{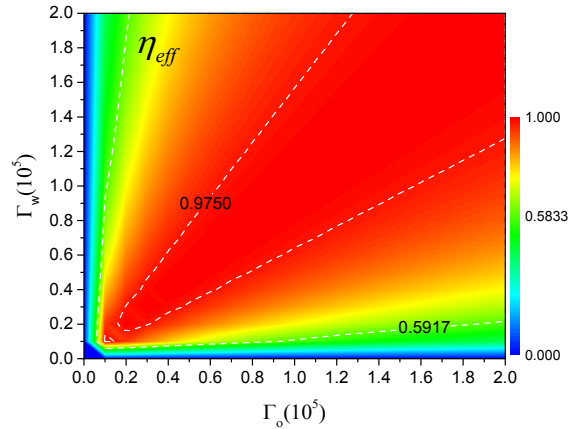}
\caption{ The effective quantum efficiency of the microwave
detector $ \eta_\mathrm{eff} $ with respect to the cooperativity
parameters $ \Gamma_\mathrm{w} $ and $ \Gamma_\mathrm{o} $. The
other parameters are the same as Fig.~\ref{fig2}.  }
\label{fig4}
\end{figure}

From Eq.~(\ref{efficency}) we can see that the mean number of
detected optical photons has a term $\eta_\mathrm{eff} n_p$ which
is proportional to the mean number of photons which were present
in the input microwave pulse. This is equivalent to having a
beam-splitter with transmissivity $\eta_\mathrm{eff}$ mixing the
incoming microwave field with vacuum fluctuations. Then there is
an additional term $\eta N_{\mathrm{thermal}}$ which accounts for
the Markovian thermalization occurring in the conversion process,
and depending on the thermal baths of the mechanical oscillator
and intracavity microwave field.

Working at cryogenic temperatures (e.g., $4$K), the thermalization
term $N_{\mathrm{thermal}}$ can be neglected and the effective
quantum efficiency of the microwave detector is indeed
$\eta_\mathrm{eff}= \eta |B(0)|^2$, where $B(0)$ accounts for the
strength of the coupling between the optical and microwave fields
with the mechanical resonator, which is monotonically increasing
in the two cooperativity terms. Assuming a high-efficient optical
detector $\eta \simeq 1$ and other achievable experimental
conditions, the quantum efficiency $\eta_\mathrm{eff}$ can
approach 100$\%$ for sufficiently high values of the cooperativity
parameters, as shown in Fig.~\ref{fig4}.

\textit{Conclusions.} We have designed a model of microwave
photodetector whose working mechanism is based on the use of an
electro-opto-mechanical converter. In our approach, microwave
fields with a small number of photons can efficiently be converted
into optical fields, which are then subject to standard optical
measurements. We have shown that overall quantum efficiency of our
detector can be very high assuming cryogenic temperatures and
achievable experimental parameters. Our receiver can potentially
be used in all those scenarios connected with the detection of
faint microwave signals, including deep space communications and
radio astronomy, for instance, in the mapping of the cosmic
background radiation.

\textit{Acknowledgments.} S.B. is grateful for support from the
Alexander von Humboldt foundation. M.C.O. acknowledges support
from CNPq/FAPESP through the Instituto Nacional de Ci\^encia e
Tecnologia em Informa\c c\~ao Qu\^antica (INCT-IQ) and FAPESP
through the Research Center in Optics and Photonics (CePOF). S.P.
has been supported by a Leverhulme Trust research fellowship and
EPSRC (via \lq qDATA\rq , grant no. EP/L011298/1).

\textit{Note added.} Upon completion of our manuscript, we noted
that a similar work by Zhang \textit{et al.} has been recently
submitted to the arxiv (arXiv:1410.0070).


\begin{thebibliography}{10}
\bibitem{langevin}The dynamics of the system can be described by following quantum Langevin equations
\begin{eqnarray}\label{qles2}
\hat{\dot{c}}_{\mathrm{w}}&&=-\kappa_{\mathrm{w}} \hat c_{\mathrm{w}}-iG_{\mathrm{w}} \hat b+\sqrt{2\kappa_{\mathrm{w}}^{\mathrm{ext}}}\hat c_{\mathrm{w,ext}}+\sqrt{2\kappa_{\mathrm{w}}^{\mathrm{int}}}\hat c_{\mathrm{w,int}},\nonumber\\
\hat{\dot{c}}_{\mathrm{o}}&&=-\kappa_{\mathrm{o}} \hat c_{\mathrm{o}}-iG_{\mathrm{o}} \hat b +\sqrt{2\kappa_{\mathrm{o}}^{\mathrm{ext}}}\hat c_{\mathrm{o,ext}}+\sqrt{2\kappa_{\mathrm{o}}^{\mathrm{int}}}\hat c_{\mathrm{o,int}}\nonumber\\
\hat{\dot{b}}&&=-\gamma_M \hat b-iG_{\mathrm{o}} \hat c_{\mathrm{o}}-iG_{\mathrm{w}} \hat c_{\mathrm{w}}+\sqrt{2\gamma_M}\hat b_{\mathrm{ in}}.\nonumber
\end{eqnarray}
where $\hat b_{in} $ and  $\hat c_j,_{\mathrm{int}}$ are the
quantum noise operators of the mechanical resonator and cavity $ j
$, respectively, while $\hat c_{j,\mathrm{ ext}}$ describes the
signal fluctuations at the input of cavity $j$. The noise
operators satisfy white-noise correlation functions $\langle \hat
b_{in}(t) \hat b_{in} ^{\dagger}(t')\rangle = \langle \hat
b_{in}^{\dagger}(t) \hat b_{in}(t')\rangle
+\delta(t-t')=(\bar{n}_b^T+1)\delta(t-t')$ and $ \langle
c_{j,\mathrm{int}}(t) c_{j,\mathrm{int}} ^{\dagger}(t')\rangle =
\langle c_{j,\mathrm{int}} ^{\dagger}(t)
c_{j,\mathrm{int}}(t')\rangle
+\delta(t-t')=(\bar{n}_j^T+1)\delta(t-t')$.
\bibitem{parameters}The explicit expressions of the coefficients are \begin{eqnarray}\label{coef}
A(\omega)&&=\frac{[\tilde{\omega}_{\mathrm{o}}-2\kappa_{\mathrm{o}}^{\mathrm{ext}}/\kappa_\mathrm{o}][\Gamma_{\mathrm{w}}+\tilde{\omega}_{\mathrm{w}}\tilde{\omega}_b]+\Gamma_{\mathrm{o}}\tilde{\omega}_{\mathrm{w}}}{\tilde{\omega}_{\mathrm{w}}[\tilde{\omega}_{\mathrm{o}}\tilde{\omega}_b+\Gamma_{\mathrm{o}}]+\Gamma_{\mathrm{w}}\tilde{\omega}_{\mathrm{o}}},\nonumber\\
B(\omega)&&=2\sqrt{\frac{\kappa_{\mathrm{o}}^{\mathrm{ext}}}{\kappa_{\mathrm{o}}}}\sqrt{\frac{\kappa_{\mathrm{w}}^{\mathrm{ext}}}{\kappa_{\mathrm{w}}}}\frac{\sqrt{\Gamma_{\mathrm{w}}\Gamma_{\mathrm{o}}}}{\tilde{\omega}_{\mathrm{w}}[\tilde{\omega}_{\mathrm{o}}\tilde{\omega}_b+\Gamma_{\mathrm{o}}]+\Gamma_{\mathrm{w}}\tilde{\omega}_{\mathrm{o}}},\nonumber\\
C(\omega)&&=\sqrt{\frac{\kappa_{\mathrm{o}}^{\mathrm{ext}}}{\kappa_{\mathrm{o}}}}\frac{2\mathrm{i}\sqrt{\Gamma_{\mathrm{o}}}\tilde{\omega}_{\mathrm{w}}}{\tilde{\omega}_{\mathrm{w}}[\tilde{\omega}_{\mathrm{o}}\tilde{\omega}_b+\Gamma_{\mathrm{o}}]+\Gamma_{\mathrm{w}}\tilde{\omega}_{\mathrm{o}}},\nonumber\\
D(\omega)&&=2\sqrt{\frac{\kappa_{\mathrm{o}}^{\mathrm{ext}}}{\kappa_{\mathrm{o}}}}\sqrt{\frac{\kappa_{\mathrm{w}}^{\mathrm{int}}}{\kappa_{\mathrm{w}}}}\frac{\sqrt{\Gamma_{\mathrm{o}}\Gamma_{\mathrm{w}}}}{\tilde{\omega}_{\mathrm{w}}[\tilde{\omega}_{\mathrm{o}}\tilde{\omega}_b+\Gamma_{\mathrm{o}}]+\Gamma_{\mathrm{w}}\tilde{\omega}_{\mathrm{o}}},\nonumber\\
E(\omega)&&=2\sqrt{\frac{\kappa_{\mathrm{o}}^{\mathrm{int}}}{\kappa_{\mathrm{o}}}}\sqrt{\frac{\kappa_{\mathrm{o}}^{\mathrm{ext}}}{\kappa_{\mathrm{o}}}}\frac{\Gamma_{\mathrm{w}}+\tilde{\omega}_{\mathrm{w}}\tilde{\omega}_b}{\tilde{\omega}_{\mathrm{w}}[\tilde{\omega}_{\mathrm{o}}\tilde{\omega}_b+\Gamma_{\mathrm{o}}]+\Gamma_{\mathrm{w}}\tilde{\omega}_{\mathrm{o}}},\nonumber
\end{eqnarray}
where $ \tilde \omega_j=1-i\frac{\omega}{\kappa_j} $ ($ j=\mathrm{w,\,o} $) and  $\tilde \omega_b=1-i\frac{\omega}{\gamma_M} $.
\end{thebibliography}
\end{document}